\documentclass[%
 aip,
 jap,
preprint,
floatfix,
]{revtex4-1}

\usepackage{setspace}
\usepackage{amsmath}
\usepackage{bm}
\usepackage{graphicx}
\begin{document}
\title{Electron-Phonon Coupling and Thermal Conductance at a Metal-Semiconductor Interface: First-principles Analysis}
\author{Sridhar Sadasivam}
\affiliation{Department of Mechanical Engineering and Birck Nanotechnology Center, Purdue University, West Lafayette, IN 47907}
\author{Umesh V. Waghmare}
\affiliation{Theoretical Sciences Unit, Jawaharlal Nehru Centre for Advanced Scientific Research, Jakkur, Bangalore 560064}
\author{Timothy S. Fisher}
\email{tsfisher@purdue.edu}
\affiliation{Department of Mechanical Engineering and Birck Nanotechnology Center, Purdue University, West Lafayette, IN 47907}

\begin{abstract}
The mechanism of heat transfer and the contribution of electron-phonon coupling to thermal conductance of a metal-semiconductor interface remains unclear in the present literature. We report ab initio simulations of a technologically important titanium silicide (metal) - silicon (semiconductor) interface to estimate the Schottky barrier height (SBH), and the strength of electron-phonon and phonon-phonon heat transfer across the interface. The electron and phonon dispersion relations of TiSi$_2$ with C49 structure and the TiSi$_2$-Si interface are obtained using first-principles calculations within the density functional theory (DFT) framework. These are used to estimate electron-phonon linewidths and the associated Eliashberg function that quantifies coupling. We show that the coupling strength of electrons with interfacial phonon modes is of the same order of magnitude as coupling of electrons to phonon modes in the bulk metal, and its contribution to electron-phonon interfacial conductance is comparable to the harmonic phonon-phonon conductance across the interface. 
\end{abstract}
\maketitle
\section{Introduction}
Thermal transport considerations form an increasingly important element in the design of miniaturized electronic devices. All semiconductor devices possess metal contacts, and hence the study of transport through metal-semiconductor interfaces is a technologically relevant problem. The study of fundamental mechanisms of heat transfer across a metal-semiconductor interface has received significant attention over the last two decades \cite{Stoner_prl_1992,Stoner_prb_1993,Stevens_jht_2005}. Stoner and Maris \cite{Stoner_prl_1992,Stoner_prb_1993} measured interfacial thermal conductance between diamond and several metals (Ti, Au, Al, and Pb) using picosecond optical techniques. They reported an anomalously high thermal interfacial conductance that exceeded the maximum thermal conductance possible from harmonic phonon transmission across the interface. Stevens et al. \cite{Stevens_jht_2005} measured the thermal conductance of a series of metal-dielectric interfaces involving Cr, Al, Au and Pt on Si, sapphire, GaN and AlN substrates. The experimental measurements were compared with predictions of the diffuse mismatch model (DMM) \cite{Swartz_modphys_1989}, which under estimated thermal conductance for impedance-mismatched interfaces while it over estimated the thermal conductance of well matched interfaces. 

\begin{figure*}[ht]
\begin{center}
\includegraphics[width=120mm,height=80mm]{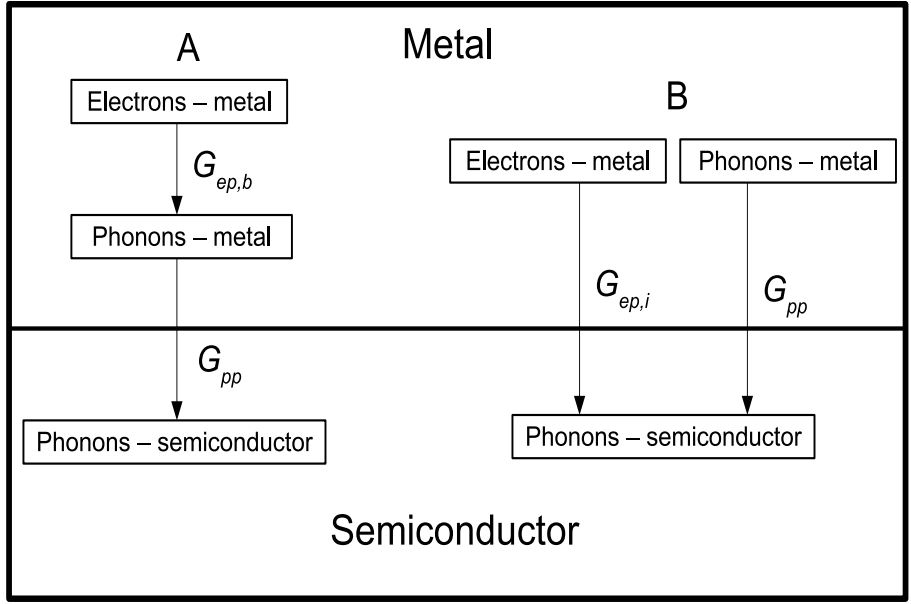}
\caption{Two mechanisms of heat transfer across a metal-semiconductor interface. Mechanism A is a two-step energy cascade where electrons in metal transfer energy to phonons in the metal and only phonon-phonon energy transfer occurs across the interface. In Mechanism B, electrons in metal couple directly with lattice vibrations in the semiconductor and this occurs in parallel with the phonon-phonon energy transfer across the interface.}
\label{mechanism_schematic}
\end{center}
\end{figure*}

Electrons are the primary heat carriers in a metal while phonons or lattice vibrations are the primary carriers in a semiconductor. At the interface between a metal and a semiconductor, an energy transfer channel from electrons in the metal to phonons in the semiconductor must exist. Numerous theoretical models have been developed \cite{Majumdar_apl_2004,Sinha_apl_2013,Huberman_prb_1994,Sergeev_prb_1998,Sergeev_physica_1999,Mahan_prb_2009} to predict the interfacial thermal conductance of a metal-semiconductor interface. These models broadly fall into two categories as shown schematically in Figure \ref{mechanism_schematic}. In Mechanism A (see Figure \ref{mechanism_schematic}), electrons in a metal couple only with phonons in the metal, and these phonons transfer energy to phonons in the semiconductor across the interface. The total interfacial thermal resistance can be calculated using a series resistance model as explained in Refs.~\onlinecite{Majumdar_apl_2004,Sinha_apl_2013}. Majumdar and Reddy \cite{Majumdar_apl_2004} used a two-temperature model coupled with DMM to predict total interfacial thermal conductance. The volumetric electron-phonon coupling constant $G$ was estimated by fitting to experimental data, and the inclusion of electron-phonon coupling resistance to the DMM resistance improved the agreement of the theoretical model with experiments. More recently, Singh et al. \cite{Sinha_apl_2013} obtained the electron-phonon coupling constant directly from Fermi's golden rule by assuming a deformation potential for electron-phonon scattering. The magnitude of the deformation potential was obtained by fitting to the electrical resistivity of bulk metal. 

In Mechanism B (see Figure \ref{mechanism_schematic}), a direct coupling exists between electrons in the metal and phonons in the semiconductor. Huberman and Overhauser \cite{Huberman_prb_1994} were among the first to propose such a mechanism to explain the anomalous high interface conductance reported for a Pb - diamond interface \cite{Stoner_prb_1993}. They used a deformation potential model to estimate the coupling between metal electrons and interfacial phonon modes. Sergeev \cite{Sergeev_prb_1998,Sergeev_physica_1999} also modeled inelastic electron-phonon scattering at the interface using an analogy with inelastic electron-impurity scattering. More recently, Mahan \cite{Mahan_prb_2009} derived an analytical theory for predicting thermal conductance between a metal and an ionic insulator. The scattering matrix elements were derived taking into account the interaction between conduction electrons in metal and the potential generated by images of ions in the insulator.

High-fidelity techniques such as the atomistic Green's function (AGF) method and molecular dynamics exist for determining the phonon-phonon coupling resistance across an interface. However, models for electron-phonon coupling near a metal-semiconductor interface assume a particular mechanism of transport (Mechanism A or B as shown in Figure \ref{mechanism_schematic}) and derive interface conductances using deformation potentials as empirical parameters. Such models offer insights into the various heat transfer mechanisms associated with electron-phonon coupling, but they do not possess the fidelity to assess the different heat transfer pathways shown in Figure 1 quantitatively. A unified approach to obtain a quantitative estimate of the thermal conductances associated with the various energy transfer pathways (electron-phonon and phonon-phonon) in Figure \ref{mechanism_schematic} is missing in the existing literature. 

In this paper, the electron-phonon coupling strength is determined directly from first-principles density functional theory (DFT) calculations. We obtain the strength of electron-phonon coupling for a bulk metal and for a metal-semiconductor interface using electron-phonon coupling linewidths derived from first-principles calculations. Relevant limits of phonon transport across the interface are treated in a simplified manner using the radiation limit and diffuse mismatch models. The radiation limit gives an upper limit of harmonic phonon interface conductance and enables us to assess the contribution of electron-phonon coupling relative to the contribution of phonons to the interface conductance. 

Transition metal silicides are widely used as metal contacts in silicon integrated circuits; titanium, nickel, cobalt and palladium silicides are common choices \cite{Tung_solid_films_1982}. We chose to study titanium silicide (TiSi$_2$) because of its technological importance in VLSI applications \cite{Alperin_IEEE_SSC_1985}. Titanium silicide (TiSi$_2$) exists in three phases: an orthorhombic base centered C49 phase, an orthorhombic face centered C54 phase and a hexagonal C40 phase \cite{Mattheiss_prb_1989,Colinet_apl_2005}. TiSi$_2$ is experimentally synthesized by evaporating a thin layer of Ti on Si followed by annealing. The annealing temperature determines the structure of TiSi$_2$ phase; the C49 phase occurs when annealed at relatively low temperatures of about 700-870 K while the C54 phase forms at temperatures beyond 900 K \cite{Jeon_jap_1992}. In the present work, a C49 titanium silicide (metal)- silicon (semiconductor) interface is chosen as a model metal-semiconductor interface to analyze thermal transport using first-principles DFT calculations. 

The remainder of the paper is organized as follows: we describe computational details in Section \ref{comp_details} followed by a brief description of the mathematical formulation of electron-phonon coupling in Section \ref{e_ph_coupling}. In Section \ref{dft_results}, we present results from DFT calculations on bulk TiSi$_2$ and a TiSi$_2$-Si interface. Complete spectral dispersion of electrons and phonons along with results for electron-phonon coupling are presented in Section \ref{dft_results}. In Section \ref{thermal_modeling}, we analyse the DFT results and estimate electron-phonon thermal conductances, and compare them with interfacial phonon conductance obtained from simplified models. 
\section{Computational Methods}\label{comp_details}
All first-principles calculations in this paper are based on the DFT framework as implemented in Quantum Espresso \cite{Giannozzi_JPhysics_2009}, which uses a planewave basis set and an ultrasoft pseudopotential to model the potential of nucleus and core electrons. The exchange correlation energy is approximated with a generalized gradient approximation (GGA) using the Perdew-Wang 91 functional form, which has been used and well-tested in prior work on TiSi$_2$ \cite{Via_apl_2001}. An energy cutoff of 680 eV (50 Ry) is used to truncate the plane-wave expansion of the wavefunctions, and an energy cutoff of 6800 eV (500 Ry) is imposed on the plane-wave basis used to represent charge density. The Brillouin zone is sampled with a uniform $10\times10\times10$ grid of k-points in calculations (for the structural relaxation) of bulk TiSi$_2$, and a grid of $10\times10\times1$ is used in the interface calculations. The cutoff energies and k-point grids are chosen to ensure that further increase in cutoff energies or refinement of k-point grids changed the total energy of the bulk and interface structures by less than 10 meV. 

A full relaxation (of both internal atomic positions and cell parameters) is performed for the bulk TiSi$_2$ calculations while a more restricted relaxation (see Section \ref{interface}) is performed for the TiSi$_2$-Si supercell. The relaxation process terminates when the Hellmann-Feynman forces on each atom decrease below 0.025 eV/$\text{\AA}$, and when the change in energy between successive relaxation iterations is less than 1 meV. The band alignment and Schottky barrier height (SBH) are estimated using the electrostatic potential method and the projected density of states method. Both methods are found to give similar estimates of the SBH. The dynamical matrices (and phonons) are obtained using density functional perturbation theory (DFPT) for which the dynamical matrices are computed on a $2\times 2\times 2$ grid for bulk C49 TiSi$_2$ and on a $2\times 2\times 1$ grid for the supercell consisting of the TiSi$_2$-Si interface. Using this approach with Fourier interpolation, dynamical matrices are obtained at arbitrary q-vectors.

\section{Mathematical Formulation of Electron-Phonon Coupling}\label{e_ph_coupling}
The definitions of electron-phonon coupling matrix elements, phonon linewidth and the associated Eliashberg function are reviewed below for the sake of completeness and clarity, and are discussed in Refs.~\onlinecite{Ziman_1960,Allen_prl_1987,Bauer_prb_1998} in more detail. The coupling matrix element $g_{\bm{k}\nu,\bm{k}+\bm{q}\nu'}^{\bm{q}p}$ for the scattering of an electron with wavevector $\bm{k}$ in the band $\nu$ by a phonon of wavevector $\bm{q}$ and polarization $p$ to a state with wavevector $\bm{k}+\bm{q}$ in band $\nu'$ is given by:
\begin{equation}
\label{coupling_matrix}
g_{\bm{k}\nu,\bm{k}+\bm{q}\nu'}^{\bm{q}p} = \sqrt{\frac{\hbar}{2M\omega_{\bm{q}p}}}\langle\psi_{\bm{k}\nu}|\phi_{\bm{q}p}\cdot\nabla V_{scf} |\psi_{\bm{k}+\bm{q}\nu'}\rangle
\end{equation}
where $\psi$ denotes the electron wavefunction and $\phi$ denotes the phonon eigenvector. The matrix element of electron-phonon coupling is given by the dot product between the phonon eigenvector and gradient of the self-consistent potential $V_{scf}$ with respect to atomic displacements. $\omega_{\bm{q}p}$ is the phonon frequency, and $M$ denotes the mass of atom (we first consider a case of mono-atomic unit cells). The formula for multi-atom unit cells involves a summation over atoms in the unit cell, and is given in Ref.~\onlinecite{Ziman_1960}. The phonon linewidth (inverse of relaxation time) $\gamma_{\bm{q}p}$ due to electron-phonon scattering is defined as:
\begin{equation}
\label{linewidth}
\gamma_{\bm{q}p} = 2\pi\omega_{\bm{q}p}\sum\limits_{\nu\nu'}\int\frac{d^3\bm{k}}{\Omega_{BZ}}|g_{\bm{k}\nu,\bm{k}+\bm{q}\nu'}^{\bm{q}p}|^2\delta(E_{\bm{k}\nu}-E_f)\delta(E_{\bm{k}+\bm{q}\nu'}-E_f)
\end{equation}
where $E_{\bm{k}\nu}$ denotes the energy of electron in state $\bm{k}\nu$, $E_f$ is the Fermi energy and $\Omega_{BZ}$ is the volume of the Brillouin zone. The foregoing expression for phonon linewidth is valid for low temperatures at which electron-phonon interactions are confined to a narrow region around the Fermi surface (see Ref.~\onlinecite{Allen_prb_1972}). The electron wavefunctions used to compute the coupling matrix elements are obtained on a $20\times 20\times 20$ k-point grid in bulk TiSi$_2$ and on a $20\times 20\times 1$ k-point grid for the TiSi$_2$-Si interface supercell. Such fine k-point grids are required for the convergence of the delta functions in Eq.~(\ref{linewidth}). The Eliashberg function $\alpha^2F(\omega)$ is defined in terms of a summation of the phonon linewidths \cite{Allen_prl_1987,Bauer_prb_1998}:
\begin{equation}
\label{Eliashberg}
\alpha^2F(\omega) = \frac{1}{2\pi D(E_f)}\sum\limits_{\bm{q},p}\frac{\gamma_{\bm{q}p}}{\hbar\omega_{\bm{q}p}}\delta(\omega-\omega_{\bm{q}p})
\end{equation}
where $D(E_f)$ is the electron density of states (DOS) at the Fermi energy. We now define a \textit{local} Eliashberg function in a manner similar to the definition of the local phonon DOS:
\begin{widetext}
\begin{eqnarray}
\label{Eliashberg_local_decomp}
\alpha^2F(\omega) &=& \frac{1}{2\pi D(E_f)}\sum\limits_{\bm{q},p}\frac{\gamma_{\bm{q}p}}{\hbar\omega_{\bm{q}p}}\delta(\omega-\omega_{\bm{q}p})\nonumber \\
&=& \frac{1}{2\pi D(E_f)}\sum\limits_{\bm{q},p}\frac{\gamma_{\bm{q}p}}{\hbar\omega_{\bm{q}p}}\delta(\omega-\omega_{\bm{q}p})\sum\limits_{l}\phi_{\bm{q}p,l}\phi_{\bm{q}p,l}^* \qquad \left(\sum\limits_{l}\phi_{\bm{q}p,l}\phi_{\bm{q}p,l}^*=1\right) \nonumber \\
&=& \frac{1}{2\pi D(E_f)}\sum\limits_{l}\sum\limits_{\bm{q},p}\frac{\gamma_{\bm{q}p}}{\hbar\omega_{\bm{q}p}}\delta(\omega-\omega_{\bm{q}p})\phi_{\bm{q}p,l}\phi_{\bm{q}p,l}^* \nonumber \\
&=& \sum\limits_{l} \alpha^2F_l(\omega)
\end{eqnarray}
\end{widetext}
where $l$ denotes atomic index within a unit cell. $\alpha^2F_l(\omega)$ is the local Eliashberg function projected onto the atom $l$, and the sum over all the local (atomic) Eliashberg functions gives the conventional total Eliashberg function. 

\begin{table}[ht]
\begin{center}
\caption{Comparison of equilibrium lattice parameters of TiSi$_2$ in bulk C49 structure obtained from the present calculation with first-principles and experimental data reported in prior literature. The relaxed internal atomic positions (in the basis of the primitive lattice vectors) obtained from the present calculation are Ti: ($\pm$0.1,$\pm$ 0.1,$\pm$0.25) and Si: ($\pm$0.75,$\pm$0.75,$\pm$0.25), ($\pm$0.44,$\pm$0.44,$\pm$0.25). VASP (LDA and GGA) values are obtained from Ref.~\onlinecite{Via_apl_2001}. The range of experimental data is also obtained from Ref.~\onlinecite{Via_apl_2001} and other references cited therein.}
\label{bulk_TiSi2_lat}
\begin{tabular}{|l|l|l|l|l|}
\hline
Method            & a ($\text{\AA}$) & b ($\text{\AA}$) & c ($\text{\AA}$) \\ \hline
Present (QE, GGA) & 3.55            & 13.62           & 3.58  \\ \hline
VASP, LDA         & 3.49            & 13.27           & 3.52             \\ \hline
VASP, GGA         & 3.55            & 13.62           & 3.58           \\ \hline
Experiment        & 3.55-3.62        & 13.49-13.77      & 3.55-3.65   \\ \hline    
\end{tabular}
\end{center}
\end{table}

\section{Electronic Structure and Phonons of Bulk T\lowercase{i}S\lowercase{i}$_2$ and T\lowercase{i}S\lowercase{i}$_2$-S\lowercase{i} Interface}\label{dft_results}
\subsection{Bulk C49 TiSi$_2$}\label{bulk} 
\begin{figure}
\begin{center}
\includegraphics[width=90mm,height=130mm]{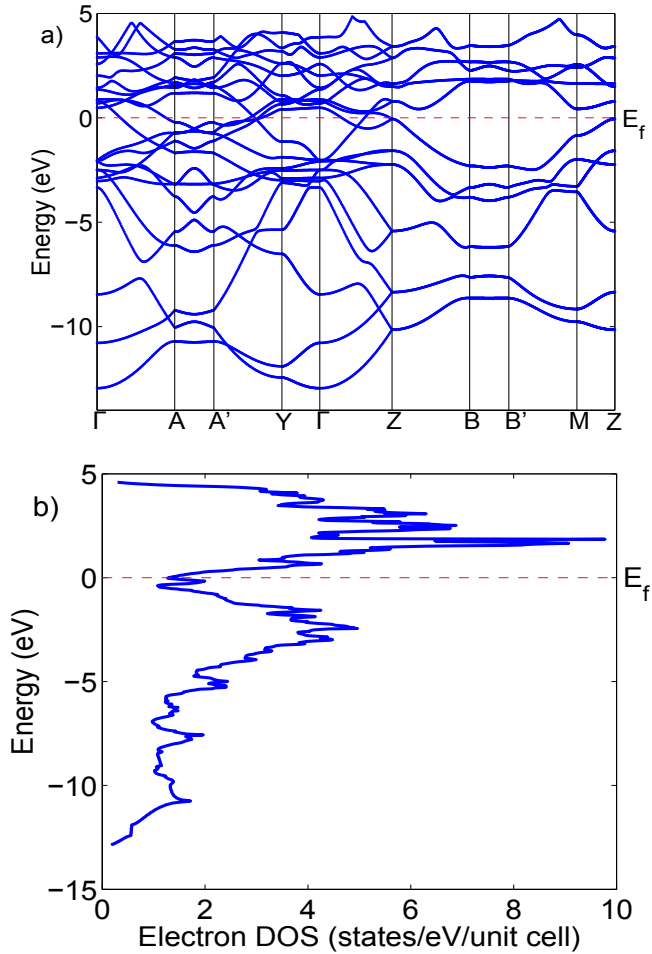}
\caption{Electron bandstructure and DOS of bulk C49 TiSi$_2$.}\label{electron_bands_dos}	
\end{center}
\end{figure}

C49 TiSi$_2$ exists in an orthorhombic structure with six atoms (2 Ti and 4 Si) per primitive unit cell, with primitive lattice vectors:
\begin{equation}
\vec{a}_1 = \frac{a}{2}\hat{i}+\frac{b}{2}\hat{j} \qquad \vec{a}_2 = -\frac{a}{2}\hat{i}+\frac{b}{2}\hat{j} \qquad \vec{a}_3 = c\hat{k}
\end{equation}
Table \ref{bulk_TiSi2_lat} shows a comparison of the relaxed lattice parameters obtained from the present calculation with experimental and first-principles calculations reported in prior literature. The present results agree well with the GGA calculations of Ref.~\onlinecite{Via_apl_2001}, and also fall within the typical DFT errors of experimental data. Our calculated electronic structure and density of states (see Figure \ref{electron_bands_dos}) agree very well with the results of linear augmented-plane-wave calculations of Mattheiss and Hensel \cite{Mattheiss_prb_1989}. 

\begin{figure}
\begin{center}
\includegraphics[width=90mm,height=140mm]{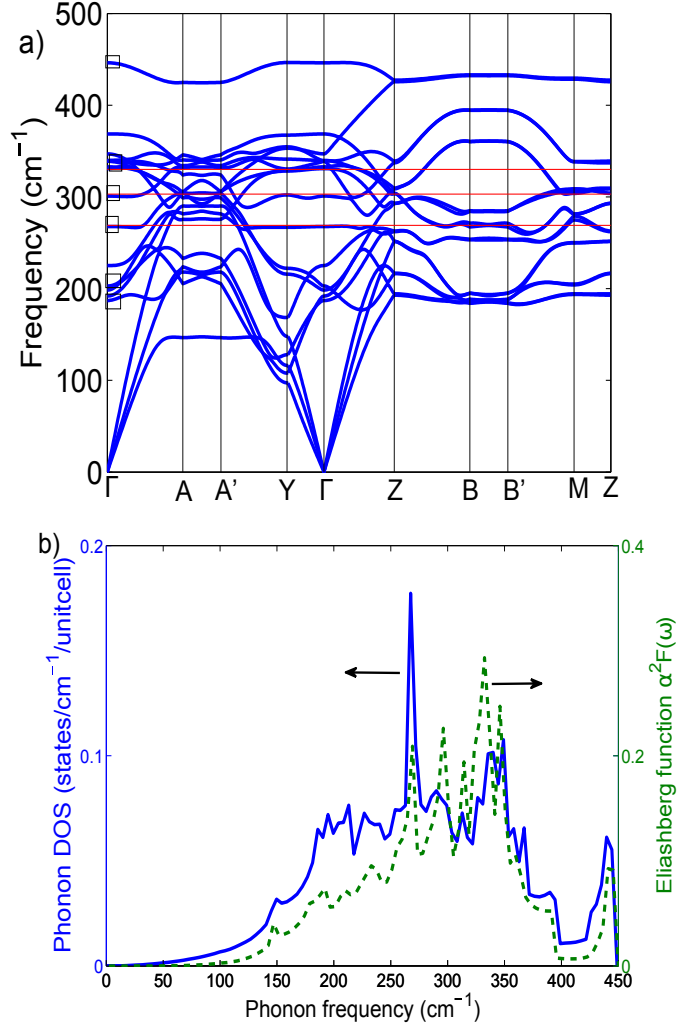}
\caption{a) Phonon dispersion of bulk C49 TiSi$_2$. The squares correspond to Raman-active modes predicted from DFPT calculations, and the horizontal lines correspond to experimental Raman frequency shifts characteristic of the C49 phase \cite{Zhao_vacuum_science_2003}. b) Phonon DOS and Eliashberg function of bulk C49 TiSi$_2$.}\label{phonon_bands_dos}
\end{center}
\end{figure}

Figure \ref{phonon_bands_dos}a shows the phonon dispersion obtained from our DFPT calculations. We are not aware of any prior reports on the phonon dispersion of C49 TiSi$_2$. However reports on the Raman characterization of TiSi$_2$ are available in the literature \cite{Jeon_jap_1992,Zhao_vacuum_science_2003}. We find excellent agreement (about 1\% error) of the DFPT results with the experimental Raman modes. These results for lattice structure and bandstructures of bulk C49 TiSi$_2$ validate the accuracy of calculational parameters, the exchange correlation energy functionals and pseudopotentials used in the present DFT formulation. 

The range of the Eliashberg function of TiSi$_2$ (see Figure \ref{phonon_bands_dos}b) is of the same order as that for typical metals such as Al, Au and Na \cite{Bauer_prb_1998}. The peaks of the Eliashberg function closely follow the peaks of the phonon DOS, as expected from the form of Eq.~(\ref{Eliashberg}). Intuitively, the Eliashberg function can be visualized as the product of phonon DOS and electron-phonon coupling linewidths. 

\subsection{C49 TiSi$_2$-Si Interface}\label{interface}
\subsubsection{Structure and Stability}
A number of experimental studies on the epitaxial growth of C49 TiSi$_2$ on Si have been reported in the literature \cite{Choi_apl_1993,Wang_jjap_1997,Tang_apl_2002,Yang_jap_2003}, however with a wide range of the Miller indices of TiSi$_2$ and Si planes that form the interface. Multiple possible epitaxial orientations have been reported within the same sample based on detailed TEM and diffraction studies of the interface. Wang et al. \cite{Wang_jjap_1997} performed a detailed study on the extent of in-plane lattice mismatch for different Miller indices and relative orientations of the TiSi$_2$ and Si planes forming the interface. Among these, the epitaxy of the (0 1 0) TiSi$_2$ plane parallel to the (0 0 1) Si plane with the [1 0 0] direction in TiSi$_2$ parallel to the [1 1 0] direction in Si had the smallest lattice mismatch in terms of superlattice area. Hence, this set of planes at the interface are likely to be the most probable in terms of energy cost of lattice mismatch and has also been reported to be the most prominent epitaxy observed \cite{Tang_apl_2002,Yang_jap_2003}. All further studies in this paper have been performed for this particular orientation at the interface. 
\begin{figure*}
\begin{center}
\includegraphics[width=140mm,height=60mm]{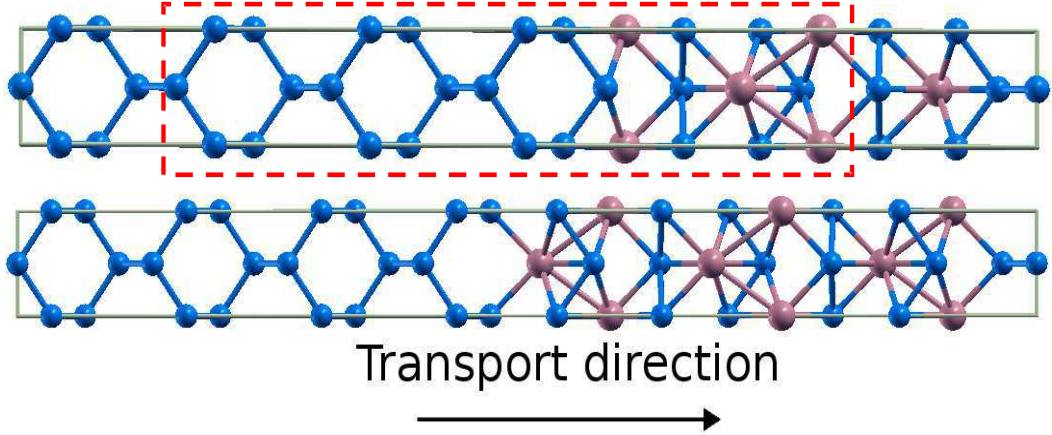}
\caption{Two interface configurations between (0 0 1) Si and (0 1 0) C49 TiSi$_2$. Blue spheres are Si atoms and magenta spheres are Ti atoms. a) Interface Configuration 1 with 28 atoms in the unit cell. The red dotted boxed region are the atomic layers considered in the calculation of Eliashberg function for interface modes in Section \ref{thermal_modeling}. b) Interface Configuration 2 with 32 atoms in the unit cell.}\label{interface_structure}	
\end{center}
\end{figure*}

In all DFT simulations of the interface, TiSi$_2$ is stretched along the [1 0 0] and [0 0 1] directions to match the bulk lattice constant of silicon. This condition mimics the experimental synthesis process where a thin film of TiSi$_2$ is typically grown on a bulk or thick Si substrate. Hence, Si can be assumed to maintain its bulk in-plane lattice constant while the thin film of TiSi$_2$ stretches in the plane of the interface. Si has a bulk lattice constant of 3.84 $\text{\AA}$ ($a_{cubic}/\sqrt{2}$) in the (0 0 1) plane while TiSi$_2$ has bulk lattice constants of 3.55 $\text{\AA}$ and 3.57 $\text{\AA}$ (see Table \ref{bulk_TiSi2_lat}) along the (0 1 0) plane. Hence a tensile strain of 7\% is required in the (0 1 0) plane of TiSi$_2$ to match the bulk lattice constant of Si. Even though this is a significant strain, we chose to work with this interface for the reasons explained above. Also, this particular epitaxial structure exhibits the smallest supercell in-plane area and hence reduces the number of atoms needed to form the supercell. This feature is important for computational efficiency in the subsequent phonon and electron-phonon coupling calculations. 

Figure \ref{interface_structure} shows unit cells of two interfacial configurations studied in the present work. Each supercell has two Si-TiSi$_2$ interfaces: one near the center and the other near the right edge of the supercell. The supercells are constructed such that both interfaces are structurally identical in the unrelaxed configuration. The atomic structures in Figure \ref{interface_structure} are relaxed by keeping the in-plane lattice constants fixed to the bulk lattice constant of Si. All internal atomic positions and the length of the unit cell along the transport (out-of-plane) direction are relaxed. A contraction in the length of the unit cell along the transport direction is observed after relaxation to minimum energy for both interfacial configurations, as expected from the tensile strain imposed on TiSi$_2$ along the in-plane direction. 

The interface energies are determined (see Table \ref{interface_properties}) using the following formula:
\begin{equation}
E_{interface} = \frac{E_{supercell}-E_{Si}^s-E_{TiSi_2}^s}{2}
\end{equation}
where $E_{supercell}$ is the energy of the supercell shown in Figure \ref{interface_structure} after relaxation. $E_{Si}^s$ is calculated by relaxation after removing all atoms belonging to TiSi$_2$ in the supercell, and $E_{TiSi_2}^s$ is calculated after removing all atoms belonging to Si (the superscript $s$ indicates a surface). The factor of two accounts for the presence of two interfaces within a supercell. Both supercells in Figure \ref{interface_structure} are constructed such that the Si plane of TiSi$_2$ forms the interface with Si. TiSi$_2$ in the C49 structure has two kinds of Si atoms \cite{Mattheiss_prb_1989}, and we obtain two configurations of the interface at the planes of these two types of Si atoms. The area normalized interfacial energy is given as $E_{interface}/A$ where $A$ is the cross-sectional area. Both interfacial configurations in Figure \ref{interface_structure} have similar interface energies of about 0.2 eV/$\text{\AA}^2$. Table \ref{interface_properties} also shows the normal stresses along the Cartesian directions after relaxation. High negative stresses along the in-plane directions ($x$ and $y$) are expected due to the tensile strain imposed by stretching the TiSi$_2$ lattice. The stress along the transport direction vanishes upon relaxation of the length of the supercell in the z-direction. 
\begin{widetext}
\begin{table}[ht]
\begin{center}
\caption{Properties of the two interface configurations shown in Figure \ref{interface_structure}.}
\label{interface_properties}
\begin{tabular}{|l|l|l|l|}
\hline
Configuration & Interface Energy (eV/$\text{\AA}^2$) & Stress (kbar)                                                          & p-type SBH (eV) \\ \hline
1                       & -0.18                       & $\sigma_{xx} = -54.2$, $\sigma_{yy} = -70.23$, $\sigma_{zz} = -0.12$   & 0.47            \\ \hline
2                       & -0.19                      & $\sigma_{xx} = -65.68$, $\sigma_{yy} = -83.35$, $\sigma_{zz} = -0.06$ & -0.29           \\ \hline
\end{tabular}
\end{center}
\end{table}
\end{widetext}

\subsubsection{Band Alignment}
\begin{figure}
\begin{center}
\includegraphics[width=90mm,height=140mm]{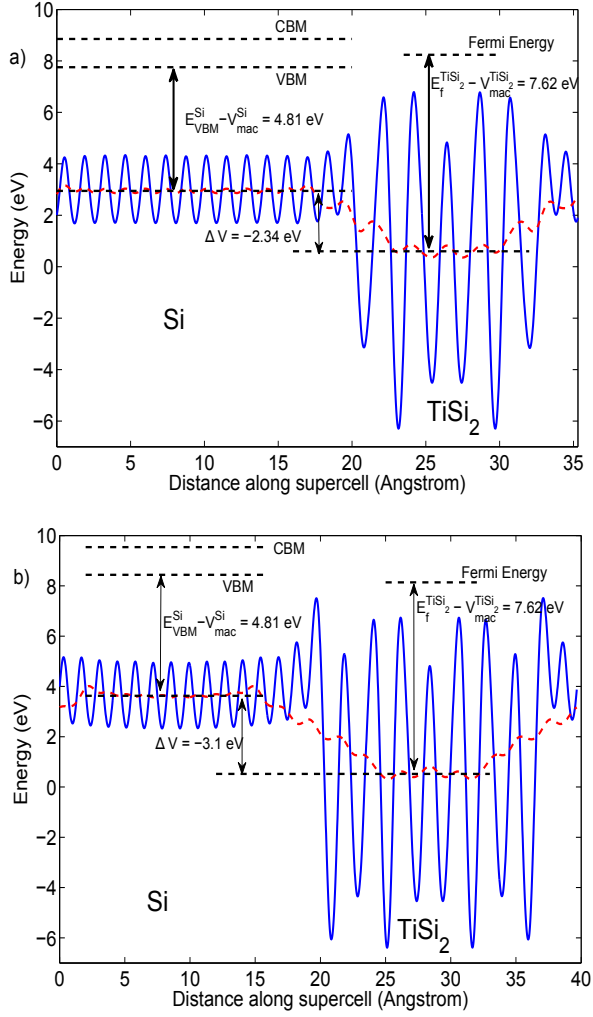}
\caption{Planar average electrostatic potential (blue continuous) and macroscopic average electrostatic potential (red dashed) plotted along the transport direction of interface Configurations 1 and 2 shown in Figure \ref{interface_structure}. The VBM in Si and Fermi energy in TiSi$_2$ are located with reference to the respective macroscopic electrostatic potentials using separate calculations on bulk Si and bulk TiSi$_2$.}\label{pot_lineup}	
\end{center}
\end{figure}

Table \ref{interface_properties} also shows the p-type SBH of the two interface configurations calculated using the band line-up method of Ref.~\onlinecite{Peressi_JPhys_1998}. According to this method, the p-type Schottky barrier $E_{SBH}$ is calculated using the following formula:
\begin{equation}
E_{SBH} = (E_{f}^{TiSi_2}-V_{mac}^{TiSi_2})-(E_{VBM}^{Si}-V_{mac}^{Si})+\Delta V
\end{equation}
where $E_f^{TiSi_2}-V_{mac}^{TiSi_2} = 7.62 \text{ eV}$ is the difference between the Fermi energy and the macroscopically averaged electrostatic potential in bulk TiSi$_2$. $E_f^{TiSi_2}-V_{mac}^{TiSi_2}$ is computed for a strained bulk TiSi$_2$ crystal with the same in-plane strain that is used to construct the lattice-matched interface. The Fermi energy in Figure \ref{pot_lineup} is located at an energy level of 7.62 eV above the macroscopically averaged electrostatic potential on the TiSi$_2$ side of the interface. The foregoing method of locating the Fermi level gives almost the same result as that obtained directly from the analysis of the self-consistent DFT solution of the relaxed supercell. $E_{VBM}^{Si}-V_{mac}^{Si} = 4.81 \text{ eV}$ is the valence band maximum (VBM) calculated with reference to the macroscopically averaged electrostatic potential in the bulk Si crystal. Although DFT is known to under-predict the band gap, the energy of VBM and its offset relative to Fermi energy are expected to be quite accurate. Hence this value is used to locate the VBM on the Si side of the interface in Figure \ref{pot_lineup}. The conduction band minimum (CBM) is located using the experimental \cite{si_bandgap} band gap (1.1 eV) of bulk Si. $\Delta V$ denotes the height of the step in the total potential at the interface and is the difference between macroscopically averaged electrostatic potentials on either side of the interface (see Figure \ref{pot_lineup}). In interface Configuration 1, the Fermi energy is found to lie between the valence and conduction bands of Si, and a p-type Schottky barrier of 0.47 eV is predicted. Our estimate of 0.47 eV is consistent with experimental SBH measurements of 0.5 - 0.6 eV in Ref.~\onlinecite{expt_sbh}. However in the interface Configuration 2, the Fermi energy lies 0.29 eV below the VBM of Si, and hence it behaves as an ohmic contact. 

\begin{figure}
\begin{center}
\includegraphics[width=90mm,height=60mm]{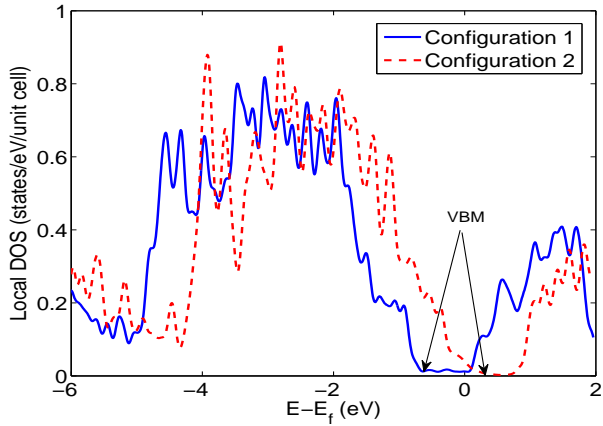}
\caption{PDOS on a Si atom farthest from the interfaces in the supercell. Configurations 1 and 2 refer to the two interface structures shown in Figure \ref{interface_structure}.}
\label{pdos_sbh}
\end{center}
\end{figure}

The estimates of SBH using the electrostatic potential method are also confirmed using the projected density of states (PDOS) method. Figure \ref{pdos_sbh} shows the PDOS on a Si atom that is farthest from the interfaces in the supercell. This PDOS is expected to mimic the bulk DOS of Si. The VBM is located about 0.6 eV below the Fermi energy of interface Configuration 1 while the VBM is located about 0.25 eV above the Fermi energy of interface Configuration 2. These values are reasonably close to the SBH estimates obtained from the electrostatic potential line-up method. The potential line-up estimates are however more reliable because the spatial convergence of local DOS to the bulk value is slower than the spatial convergence of potentials or charge density \cite{Peressi_JPhys_1998}. The foregoing results illustrate that the details of the atomic scale structure of an interface decide the behavior of a metal-semiconductor contact. 
\subsubsection{Phonons and Electron-Phonon Coupling}
\begin{figure}
\begin{center}
\includegraphics[width=90mm,height=80mm]{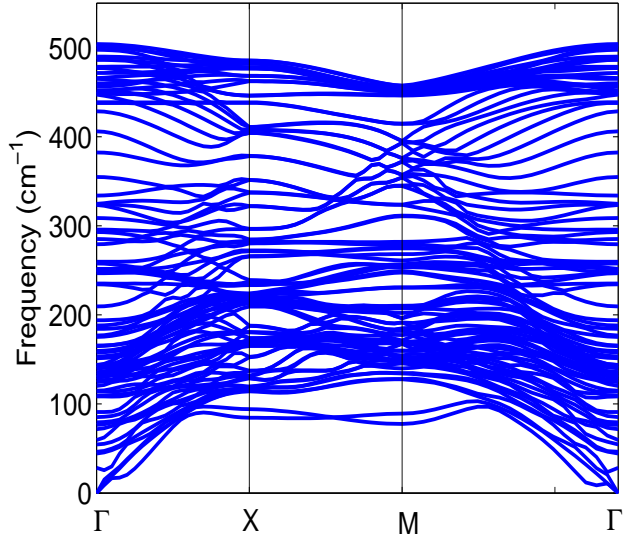}
\caption{Phonon dispersion of interface supercell Configuration 1.}
\label{interface_phonon_disp}
\end{center}
\end{figure}

DFPT calculations are computationally intensive, and hence we focus only on interface Configuration 1 (Schottky interface) for further phonon and electron-phonon coupling calculations. Figure \ref{interface_phonon_disp} shows the phonon dispersion of the supercell with TiSi$_2$-Si interface Configuration 1 calculated in the two-dimensional Brillouin zone. Real vibrational frequencies suggest that the interface configuration is stable in spite of the large tensile strain imposed on TiSi$_2$ to match its lattice with Si. We also observe the existence of many soft (low frequency and dispersionless) interfacial modes that are expected to further stabilize the structure at higher temperatures. 

\begin{figure}
\begin{center}
\includegraphics[width=100mm,height=70mm]{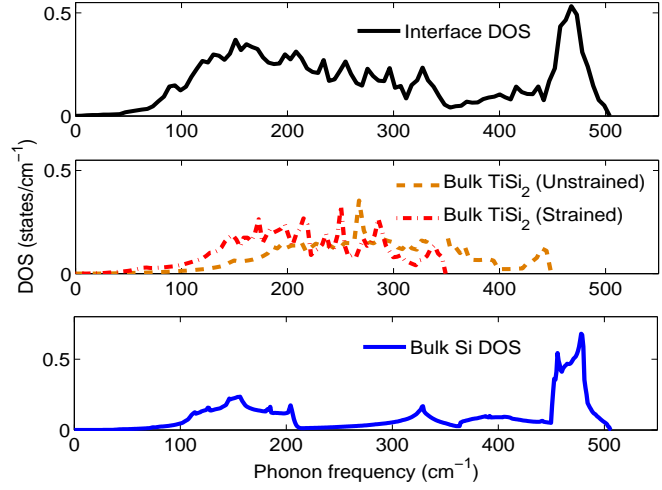}
\caption{Phonon DOS of the TiSi$_2$-Si interface supercell Configuration 1 compared with that of bulk TiSi$_2$ and bulk Si.}
\label{interface_phon_dos}
\end{center}
\end{figure}

We now examine the phonon DOS of TiSi$_2$-Si interface and compare it with that of bulk TiSi$_2$ and bulk Si (Figure \ref{interface_phon_dos}). The phonon DOS for bulk TiSi$_2$ and Si have been scaled by factors of 2 and 8 respectively because the interface supercell (see Figure \ref{interface_structure}a) contains 2 unit cells of TiSi$_2$ (4 Ti and 8 Si atoms) and 8 unit cells of Si (16 Si atoms). The bulk phonon DOS of TiSi$_2$ is plotted both for the relaxed unstrained crystal and for a structure with 7\% tensile strain. We observe that the peak DOS in bulk TiSi$_2$ (both strained and unstrained) occurs at frequencies in the range of 200-300 cm$^{-1}$ while that of bulk Si occurs in the frequency range of 150-200 cm$^{-1}$. The interfacial phonon DOS has a relatively broader peak between 150-300 cm$^{-1}$ that encompasses both these frequency ranges. This result suggests that the interfacial phonon DOS follows an average of the bulk structures forming the interface. Similar shifts in the DOS of interfacial modes have been reported for In/Si interfaces in Ref.~\onlinecite{Kaviany_prb_2010}. 

\begin{figure}
\begin{center}
\includegraphics[width=95mm,height=75mm]{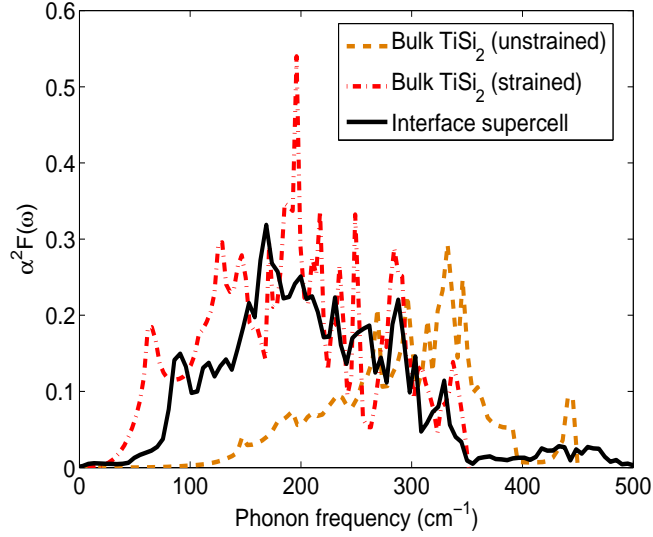}
\caption{Eliashberg function of the TiSi$_2$-Si interface Configuration 1 compared with that of bulk TiSi$_2$.}
\label{interface_Eliashberg}
\end{center}
\end{figure}

A comparison between the Eliashberg function (Figure \ref{interface_Eliashberg}) of the supercell with interface Configuration 1 with that of strained and unstrained bulk TiSi$_2$ indicates that the average Eliashberg function for the interfacial structure is of the same order of magnitude as that for bulk TiSi$_2$. Since TiSi$_2$ in the interfacial structure is strained by 7\%, Eliashberg function of the interfacial configuration follows the Eliashberg function of strained bulk TiSi$_2$. The tensile strain in the structure also produces a distinct red-shift in the low-frequency peaks of the Eliashberg function. The Eliashberg function of the strained crystal peaks in the range of 150-200 cm$^{-1}$ while that of unstrained bulk TiSi$_2$ peaks around 300 cm$^{-1}$.

%A similar red shift is also observed in the interfacial phonon DOS due to the bonding between TiSi$_2$ and Si (see Figure \ref{interface_phon_dos}). Since the Eliashberg function shows qualitatively similar spectral behavior as the phonon DOS (see Figure \ref{phonon_bands_dos}b), the red shift in the Eliashberg function of the interface supercell is attributed to the red shift in the phonon DOS of the interfacial structure.

\begin{figure}
\begin{center}
\includegraphics[width=90mm,height=130mm]{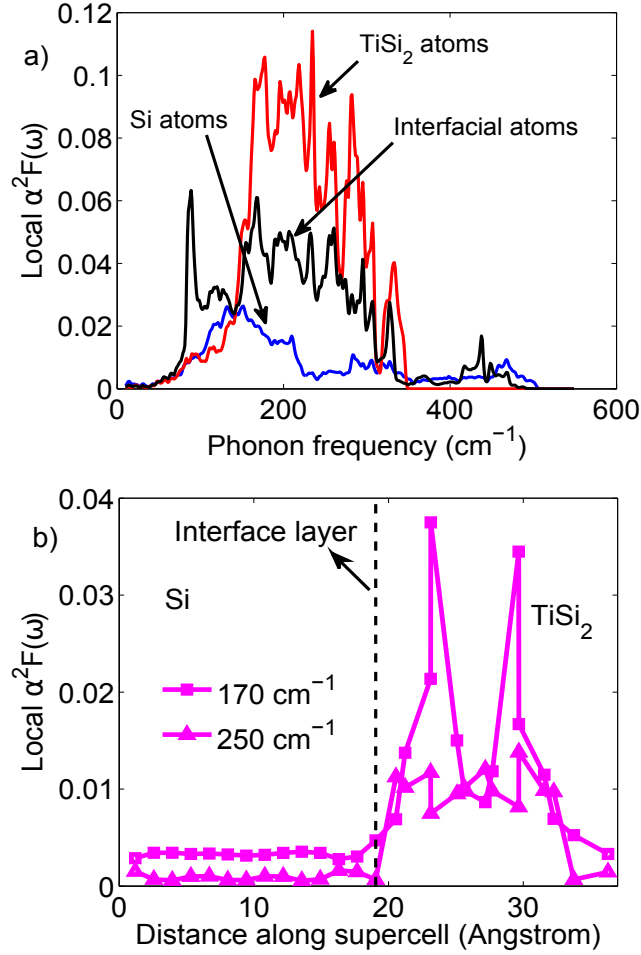}
\caption{a) Spectral variation of the Eliashberg function over three different regions of the supercell with interface Configuration 1. b) Variation of the Eliashberg function along the transport direction of the interface Configuration 1 for phonon frequencies of 170 and 250 cm$^{-1}$.}\label{local_Eliashberg}	
\end{center}
\end{figure}

The Eliashberg function can be decomposed spatially using the phonon eigenvectors as shown in Eq.~(\ref{Eliashberg_local_decomp}). Figure \ref{local_Eliashberg}a shows the spatially decomposed Eliashberg function for three different regions of the supercell: atoms belonging to Si that are farthest from the interface, atoms belonging to TiSi$_2$ that are farthest from the interface and atoms near the interfacial region. The Eliashberg function is summed over 6 atomic layers in each of these three regions. As expected, the strength of electron-phonon coupling is strongest on the metal side of the interface and weakest on the semiconductor side. The interfacial contribution to Eliashberg function is dominant in the 150-200 cm$^{-1}$ range of frequencies (the modes red-shifted due to the tensile strain). Figure \ref{local_Eliashberg}b shows the Eliashberg function projected on different atomic planes along the supercell for two specific phonons with frequencies of 170 and 250 cm$^{-1}$. The Eliashberg function is weak on the Si side but increases sharply across the interface to approximately ten times its value on the Si side. The decay in the Eliashberg function beyond layer 25 is due to the second interface (and the Si region beyond it) that exists because of the translational periodicity used in the calculations. The Eliashberg function on the Si side of the interface suggests that the phonon mode at 170 cm$^{-1}$ couples more strongly with the electronic states penetrating into the semiconductor than the mode at 250 cm$^{-1}$. As a result, this mode is expected to contribute more actively to the cross-interface electron-phonon thermal conductance. Also, the local Eliashberg function on the silicon side of the interface does not decay immediately to zero as the local electron DOS at the Fermi energy is also not exactly zero for a silicon atom farthest from the interface and the edges of the supercell (see PDOS in Figure \ref{pdos_sbh}). This is due to the slow convergence of local electron DOS to the bulk Si DOS at such small distances ($\sim$10 $\text{\AA}$) from the interface layer.

\section{Thermal Conductance of the T\lowercase{i}S\lowercase{i}$_2$-S\lowercase{i} Interface}\label{thermal_modeling}
In this section, we report predictions of thermal conductance along the different heat transfer pathways (electron-phonon, phonon-phonon) shown schematically in Figure \ref{mechanism_schematic}. The objective is to obtain a quantitative estimate of the contributions of each of these pathways, which could help assess the validity of simplified models such as those proposed in Refs.~\onlinecite{Majumdar_apl_2004,Sinha_apl_2013}. We consider three different contributions to thermal conductance (see Figure \ref{mechanism_schematic}) in this section:
\begin{itemize}
\item $G_{ep,b}$ is the electron-phonon thermal conductance in the bulk region of the metal and is obtained from the Eliashberg function of strained bulk TiSi$_2$. This conductance is used in the models of Refs.~\onlinecite{Majumdar_apl_2004,Sinha_apl_2013} (Mechanism A in Figure \ref{mechanism_schematic}).
\item $G_{ep,i}$ is the electron-phonon thermal conductance contributed by degrees of freedom near the interface region and physically represents the coupling of electrons in metal with the interfacial or joint phonon modes. $G_{ep,i}$ is obtained from the Eliashberg function of the interface supercell. This conductance was proposed to be important in the thermal conductance of a Pb/diamond interface \cite{Huberman_prb_1994} and was also considered in the works of Sergeev \cite{Sergeev_prb_1998,Sergeev_physica_1999}. Since the joint phonon modes exist on both sides of the interface, a part of the energy transferred from electrons to interfacial phonon modes is transferred across the interface. Hence, this conductance relates to direct coupling of electrons to phonons across the interface but is often neglected with the assumption that electrons in metal are thermally insulated from the semiconductor at the interface \cite{Majumdar_apl_2004,Sinha_apl_2013,Ruan_prb_2012}. 
\item $G_{pp}$ is the phonon thermal conductance of the interface and can be evaluated using advanced techniques such as the atomistic Green's function method \cite{zhang_jht_2007}. In this work, we adopt a simplified approach to estimate $G_{pp}$ using the diffuse mismatch model \cite{Swartz_modphys_1989} and also obtain an upper bound from the radiation limit.   
\end{itemize}
The energy transfer rate (per unit volume) $J_{ep}$ from electrons to phonons can be expressed in terms of the Eliashberg function using the following expression (see Ref.~\onlinecite{Allen_prl_1987} for a detailed derivation):
\begin{widetext}
\begin{equation}
J_{ep} = 2\pi D(E_f)\int\limits_{0}^{\infty}{(\hbar\omega)^2\alpha^2F(\omega)(f_{BE}^o(\omega,T_e)-f_{BE}^o(\omega,T_l))d\omega}
\end{equation}
\end{widetext}
where $T_e$ and $T_l$ denote the electron and lattice temperatures respectively, and $f_{BE}^o$ denotes the Bose-Einstein distribution function. The electron-phonon conductance $G_{ep}$ can be defined for a small difference between the electron and lattice temperatures as:
\begin{equation}
\label{Gep}
G_{ep} = 2\pi D(E_f)\int\limits_{0}^{\infty}{(\hbar\omega)^2\alpha^2F(\omega)\frac{\partial f_{BE}^o}{\partial T}d\omega}
\end{equation}
This expression is used to compute the electron-phonon conductances $G_{ep,b}$ and $G_{ep,i}$ where the first-principles Eliashberg function of bulk TiSi$_2$ and the interface supercell are used in the integrand. The latter ($G_{ep,i}$) is computed by summing the local Eliashberg function projected on 19 atomic planes around the interface near the center of the supercell. The contribution to the Eliashberg function from other layers near the edges of the supercell are not considered in order to extract only the contribution of electron-phonon coupling near a single interface. These 19 atomic layers are shown in Figure \ref{interface_structure}a and correspond to the atomic index $l$ (see Eq.~(\ref{Eliashberg_local_decomp})) in the range of 5 to 23 where $l=15$ is the location of the interface. The electron-phonon conductance computed from Eq.~(\ref{Gep}) has units of W/m$^3$K and needs to be converted to W/m$^2$K to obtain an area-normalized conductance that can be compared with $G_{pp}$. We do this by multiplying by a typical length scale for electron-phonon energy transfer. A length scale of $3\lambda_{ep}$ which corresponds to 95\% energy loss was suggested in Ref.~\onlinecite{Sinha_apl_2013} where $\lambda_{ep}$ is the electron mean free path due to scattering from phonons in bulk TiSi$_2$. Using the experimentally determined electrical resistivity ($\rho$) and carrier concentration ($n$) of C49 TiSi$_2$ in Ref.~\onlinecite{Mammoliti_jap_2002}, $\lambda_{ep}$ can be estimated using the simple formula $\lambda_{ep} = mv_f/ne^2\rho$. The low-temperature residual resistivity is subtracted from the actual resistivity at room temperature to obtain the resistivity due to electron-phonon scattering alone, excluding other scattering effects such as grain boundary and impurity scattering. The Fermi velocity of bulk C49 TiSi$_2$ is $v_f = 0.49\times10^6$ m/s \cite{Mattheiss_prb_1989}. We obtain an electron mean free path due to phonon scattering of about 4 nm at room temperature and use a length scale of 10 nm for converting the volumetric electron-phonon conductance to an area-normalized conductance.

The spatial extent of the joint phonon modes depends on the anharmonicity of the local interfacial structure and is computationally expensive to predict from first-principles calculations. A heuristic approximation for the spatial extent of joint/interfacial phonon modes was provided by Huberman and Overhauser \cite{Huberman_prb_1994} who assumed that the joint modes extend to a distance of the bulk phonon mean free path on either side of the interface. We estimate the phonon mean free path $\Lambda_p$ in bulk TiSi$_2$ using the kinetic theory expression for thermal conductivity $\kappa_p = (1/3)c_vv_g\Lambda_p$. The phonon contribution to thermal conductivity $\kappa_p$ was estimated as 3.4 W/mK in Ref.~\onlinecite{neshpor1968thermal} by subtracting the electronic thermal conductivity (obtained using the Wiedemann Franz law) from the total thermal conductivity measured experimentally. An average phonon group velocity of 5900 m/s is estimated using the phonon dispersion in Figure \ref{phonon_bands_dos}a. The specific heat capacity of C54 TiSi$_2$ was experimentally measured and reported as 66.9 J/mol K in Ref.~\onlinecite{agarwal2001heat}. We assume the same value for C49 TiSi$_2$ as the specific heat is not expected to be significantly different. This gives a phonon mean free path in C49 TiSi$_2$ as 7 $\text{\AA}$ while the phonon mean free path in silicon is about 400 $\text{\AA}$ \cite{Johnson_prl_2013}. Such a large difference in phonon mean free paths indicates that the energy of the joint modes is predominantly concentrated on the Si side of the interface. Hence the electron-joint phonon mode conductance $G_{ep,i}$ calculated using the Eliashberg function of the supercell containing the interface is expected to be a good estimate of the cross-interface electron-phonon thermal conductance. We note that this approach is similar to the Pb/diamond interface considered in Ref.~\onlinecite{Huberman_prb_1994} where the energy of the joint modes is concentrated primarily in diamond due to the small phonon mean free path in Pb. 

\begin{figure}
\begin{center}
\includegraphics[width=90mm,height=75mm]{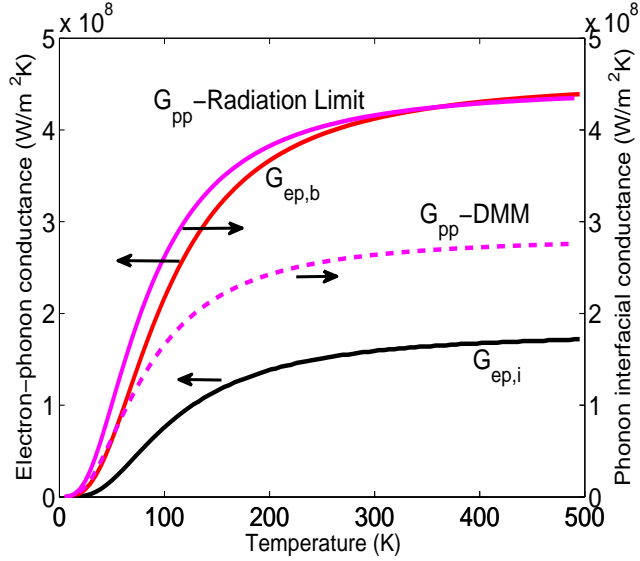}
\caption{Thermal conductance due to electron-phonon coupling for bulk TiSi$_2$ ($G_{ep,b}$) and the interface ($G_{ep,i}$) along with interfacial phonon-phonon thermal conductances from the radiation limit and DMM. The arrows indicate the y-axis to which each curve corresponds to.}
\label{thermal_cond}
\end{center}
\end{figure}

The phonon-phonon interfacial conductance $G_{pp}$ is estimated from the Landauer formula:
\begin{equation}
G_{pp} = \left(\frac{1}{2\pi}\right)^3\sum\limits_{p}\int\limits_{\bm{q},q_z>0}\hbar\omega(\bm{q},p)v_{g,z}\frac{\partial f_{BE}^o}{\partial T}\mathcal{T}(\omega(\bm{q},p))d^3\bm{q}
\end{equation}
where $v_{g,z}$ denotes the phonon group velocity along the transport direction. The transmission function $\mathcal{T}(\omega(\bm{q},p))$ is estimated in the harmonic radiation limit (upper limit for harmonic interface conductance) and also from the diffuse mismatch model (DMM) using the calculated full phonon dispersions of strained bulk TiSi$_2$ and bulk Si. Ref.~\onlinecite{Stoner_prb_1993} provides details on the radiation limit of interface conductance, and Ref.~\onlinecite{Reddy_apl_2005} describes implementation of DMM using full phonon dispersion rather than the more common Debye approximation. 

Figure \ref{thermal_cond} shows the variation of $G_{pp}$, $G_{ep,b}$ and $G_{ep,i}$ with temperature. While $G_{ep,i}$ is lower than $G_{pp}$ and $G_{ep,b}$ at room temperature it is not low enough that direct coupling from electrons in metal to interfacial phonon modes can be neglected. A model of resistors in series such as that used in Refs.~\onlinecite{Majumdar_apl_2004,Sinha_apl_2013,Ruan_prb_2012} assumes electron-phonon coupling across the interface to be insignificant compared to electron-phonon coupling in the bulk of the metal. Our results however indicate that the coupling strength of electrons to the interface phonon modes at a TiSi$_2$-Si interface is of the same order of magnitude as the electron-phonon coupling in bulk TiSi$_2$ metal. Thus, a simple series resistor model may not be a valid approximation in general to the thermal interface conductance. Also, the electron-phonon thermal conductances (both $G_{ep,b}$ and $G_{ep,i}$) are of the same order of magnitude as the phonon-phonon conductance across the TiSi$_2$-Si interface at room temperature. Moreover, we note that the electron-phonon interaction across an interface provides a direct coupling of the primary thermal carriers in the respective materials. Therefore, this coupling process should be considered a parallel path to the all-phonon cross-interface transport process. 

Finally, we note that although the results on electron-phonon coupling have been presented only for interface Configuration 1, the strength of cross-interface electron-phonon coupling is expected to be even stronger for interface Configuration 2 because the latter forms an ohmic contact and hence electrons from the metal experience no energy barrier for coupling with phonons in the semiconductor. These findings contrast with those of Ref.~\onlinecite{Sinha_apl_2013}, which concluded that electron-phonon coupling resistance can be ignored at temperatures above 200 K, albeit for different metal-semiconductor interfaces than considered in this work.

\section{Conclusions}
The contribution of electron-phonon coupling, especially the direct coupling of electrons in a metal to phonons in a semiconductor, is often neglected in the study of thermal transport across a metal-semiconductor interface. The present ab initio studies on a technologically important TiSi$_2$ (metal) - Si (semiconductor) interface however indicate that the strength of coupling between electrons and interfacial joint phonon modes is of the same order of magnitude as the strength of electron-phonon coupling in bulk metal. Further, the thermal conductance from the coupling of electrons to interfacial or bulk metal phonons is comparable to the phonon-phonon conductance across the interface. These results highlight the importance of considering the direct coupling of electrons in metal to phonons in semiconductor as an alternate pathway for interfacial thermal transport in addition to the conventional phonon (metal) - phonon (semiconductor) energy transfer pathway. 

Looking toward future developments, we first note that the non-equilibrium Green's function (NEGF) method for electron transport and the analogous atomistic Green's function method for phonon transport are well established in contemporary literature. However, reports of coupled electron-phonon transport using the Green's function method are scarce and exist only for simple systems such as 1D atom chains \cite{lu_prb_2007,zhang_JPhys_2013}. We are currently working on the development of a NEGF-based electron-phonon coupled transport solver to which the results from first principles calculations (such as inter-atomic force constants, electron-phonon coupling matrix elements and the associated Eliashberg function) are direct inputs.

\section*{Acknowledgements}
SS thanks the Indo-US Science and Technology Forum for supporting his exchange visit to JNCASR under the Joint Centre on Nanomaterials for Energy. We thank the Office of Naval Research (Award Number: N000141211006) for funding this research. We also acknowledge computational support from the Rosen Center for Advanced Computing at Purdue. All atomic structure visualizations in this paper were obtained using the XCrySDen package \cite{xcrysden}.
\bibliography{references}
%\addbibresource{references}
%\printbibliography
\end{document}